# Distributed Resources for the Earth System Grid Advanced Management (DREAM)

Luca Cinquini, Steve Petruzza, Jason Jerome Boutte, Sasha Ames, Ghaleb Abdulla, Venkatramani Balaji, Robert Ferraro, Aparna Radhakrishnan, Laura Carriere, Thomas Maxwell, Giorgio Scorzelli, Valerio Pascucci

# Accomplishments

The DREAM project was funded more than 3 years ago to design and implement a next-generation ESGF (Earth System Grid Federation [1]) architecture which would be suitable for managing and accessing data and services resources on a distributed and scalable environment. In particular, the project intended to focus on the computing and visualization capabilities of the stack, which at the time were rather primitive. At the beginning, the team had the general notion that a better ESGF architecture could be built by modularizing each component, and redefining its interaction with other components by defining and exposing a well defined API. Although this was still the high level principle that guided the work, the DREAM project was able to accomplish its goals by leveraging new practices in IT that started just about 3 or 4 years ago: the advent of containerization technologies (specifically, Docker), the development of frameworks to manage containers at scale (Docker Swarm and Kubernetes), and their application to the commercial Cloud. Thanks to these new technologies, DREAM was able to improve the ESGF architecture (including its computing and visualization services) to a level of deployability and scalability beyond the original expectations.

# ESGF Node Containerized Architecture

Arguably, the most important accomplishment of the DREAM project was the design and implementation of a container-based architecture for the full ESGF Node, as this deliverable fulfills the original overarching goal of the project for modularity and scaling of the ESGF software stack. This milestone was achieved over the course of two years, in collaboration with other ESGF institutions, through a process that decomposed the full stack into independent services, identified and isolated their configurations, and encapsulated each service into a portable Docker container. The final result is a set of containerized services which can be deployed either on a single server (such as a developer's laptop) via *docker-compose*, or on a cluster of nodes via *Kubernetes* and *Helm*.

The first stable release of the ESGF/Docker stack took place in September 2018. It was a well tested and stable release, although not feature complete, as it was noticeably missing any Globus functionality. Figure 1 below shows the general architecture, which relies on Kubernetes *Pods* to deploy the Docker containers, Kubernetes *Services* to enable inter-container communication, as well as access by clients, and Kubernetes Persistent Volumes to store non-ephemeral data.

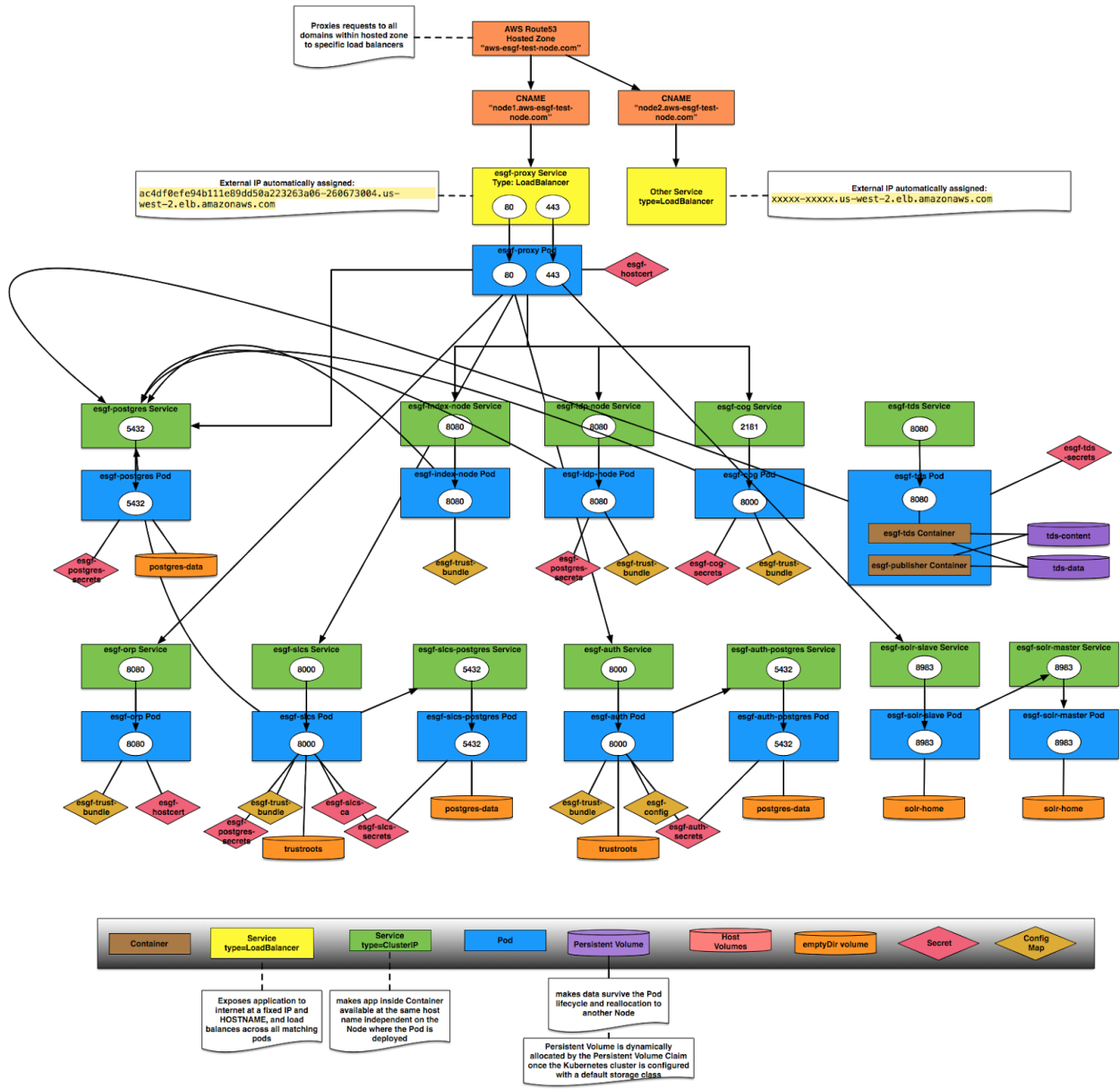

*Figure 1: ESGF/Docker software stack deployed as Kubernetes objects.*

This release was installed and tested both on a Kubernetes cluster running on the JPL internal network, and on the Amazon and Google commercial clouds (see Figure 2 below). It was also the basis of the ESGF/Pangeo testbed described later in this section.

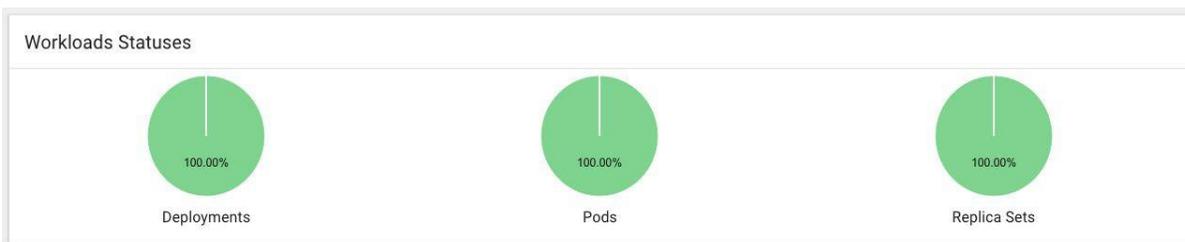

*Figure 2: Snapshots of the Kubernetes dashboard showing the ESGF/Docker software stack deployed on AWS (upper frame) and GCP (lower frame).*

In the next year, before the conclusion of the project, the DREAM project plans to issue new releases of the ESGF/Docker stack, which will make it feature complete (i.e. include

Globus), and be completely up to date with respect of the service version and configuration as installed by the classic ESGF installer. The goal is to make ESGF/Docker a viable (and preferable) alternative to the traditional method of installing and operating an ESGF Node. Another goal for the coming year is the ability to deploy only part of the stack, allowing an institution or project to tailor its deployment to its specific needs (e.g., only an Index Node, a Data Node, and Identity Provider).

# ESGF Compute Node

In 2018 the Compute Node architecture has continued to evolve, from its beginnings as a simple web application to a complex compute infrastructure (see Figure 3). We're leveraging many modern technologies like NGINx, Traefik, Celery, Redis, Postgres and more to provide a powerful and flexible system for users. Each piece of the infrastructure has been containerized providing portability and reproducibility to the software. Utilizing containers and Kubernetes, we're able to provide a scalable application that is highly available. We've designed the software architecture in a way that allows us to support multiple computational backends through a single application. We provide access to CDAT from LLNL [2], EDAS [3] from NASA and Ophidia [4] from CMCC and plan to support many more frameworks in the future. The software provides many basic building blocks such as aggregation, subsetting, regridding, min, max and average from different frameworks. Theses processes can be combined into workflows to create much more complex processes.

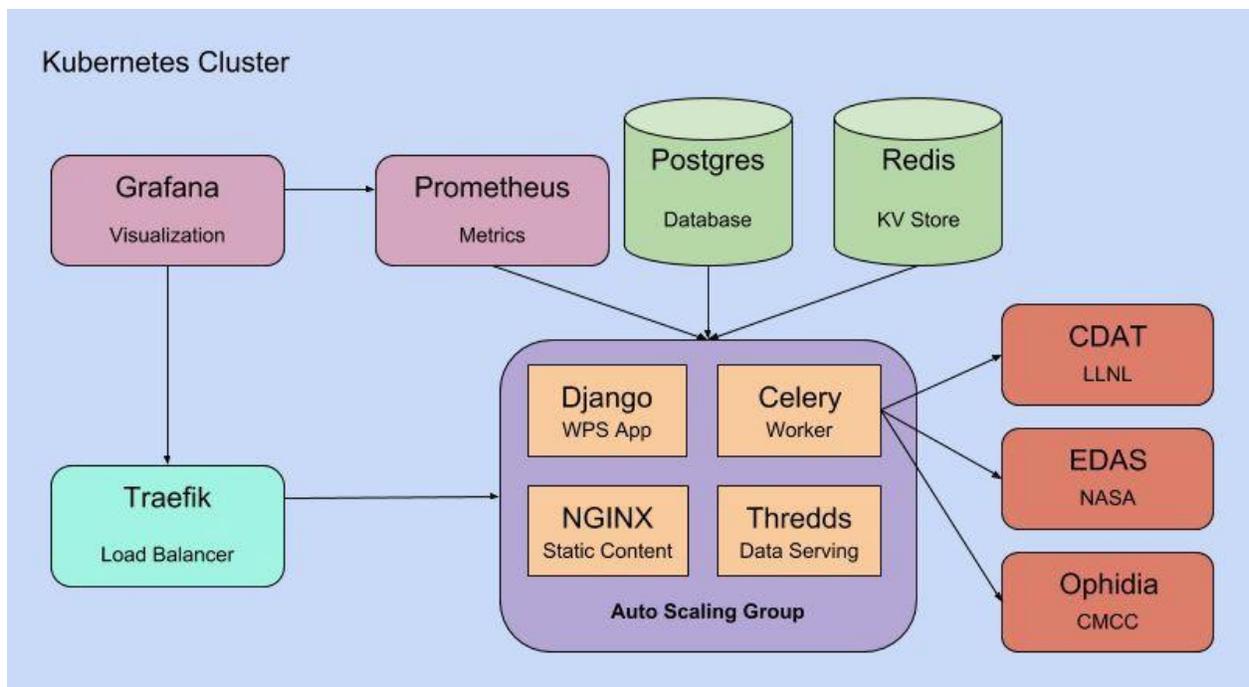

*Figure 3: Schematic architecture of the ESGF Compute Node.*

This year we've had our initial beta release in February 2018, our first stable release in April and a second release just this December. We plan to keep evolving this infrastructure to continue providing the community with a stable and powerful remote computational solution.

## EDAS Analytics

In 2018 NASA continued development of its Earth Data Analytic Services (EDAS) Framework, using Xarray as a high level data model, and Dask for distributing the analysis on all available computing resources. This framework enables scientists to execute data processing workflows combining common analysis and forecast operations, with direct access to the massive data stores at NASA or remote access to ESGF data. The data is accessed in standard (NetCDF, HDF, etc.) formats and processed using vetted tools of earth data science, e.g. ESMF, CDAT, NCO, Keras, Tensorflow, etc. EDAS services are accessed via the ESGF Compute Working Team's WPS API. Client packages in Python, Java/Scala, or JavaScript contain everything needed to build and submit EDAS requests.

EDAS facilitates the construction of high performance parallel workflows by combining canonical analytic operations to enable processing of huge datasets within limited memory spaces with interactive response times. EDAS services include configurable high performance machine learning modules designed to operate on the products of EDAS workflows. The EDAS architecture brings together the tools, data storage, and high performance computing required for timely analysis of large-scale data sets, where the data resides, to ultimately produce societal benefits (see Figures 4, 5).

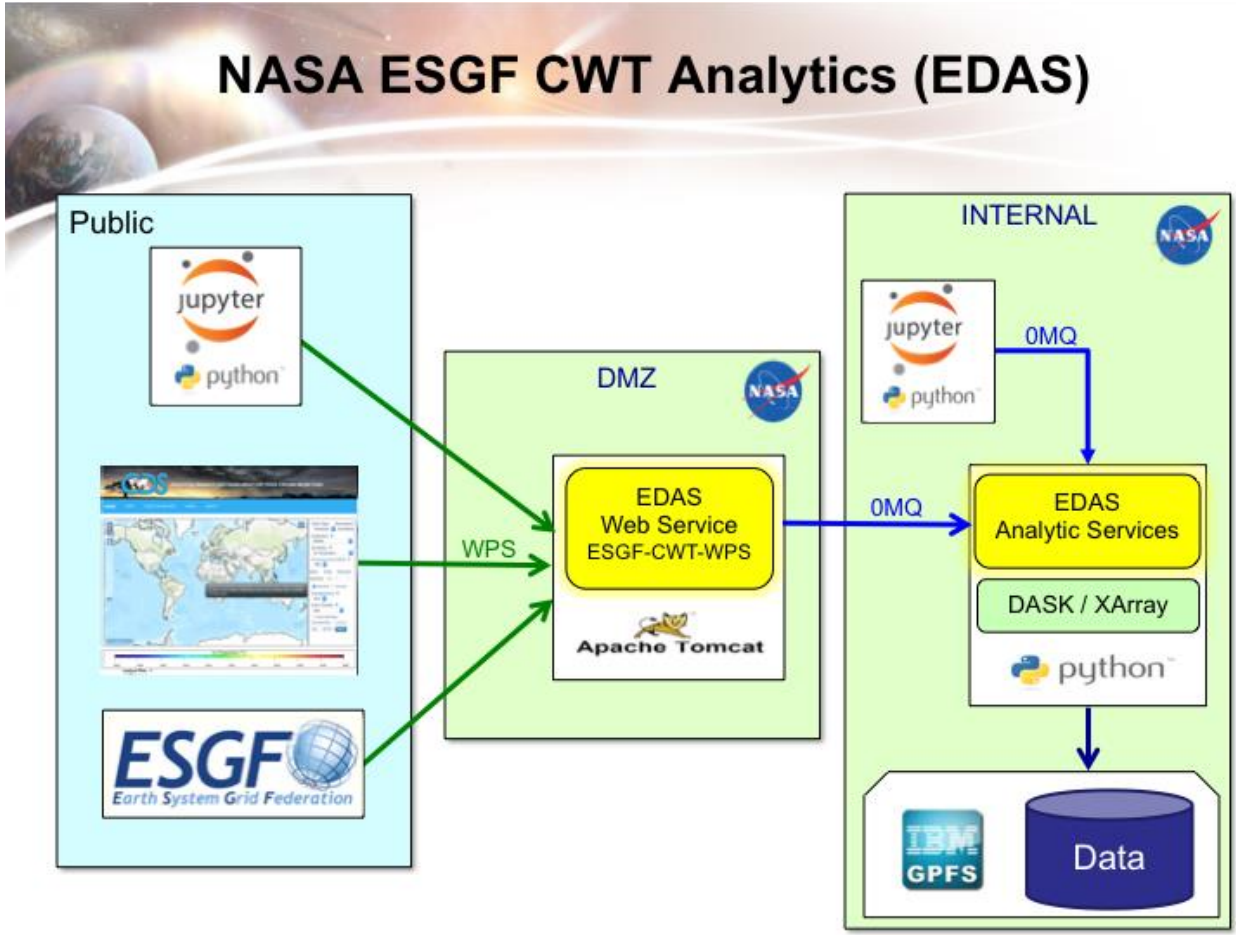

*Figure 4: EDAS Schematic Architecture*

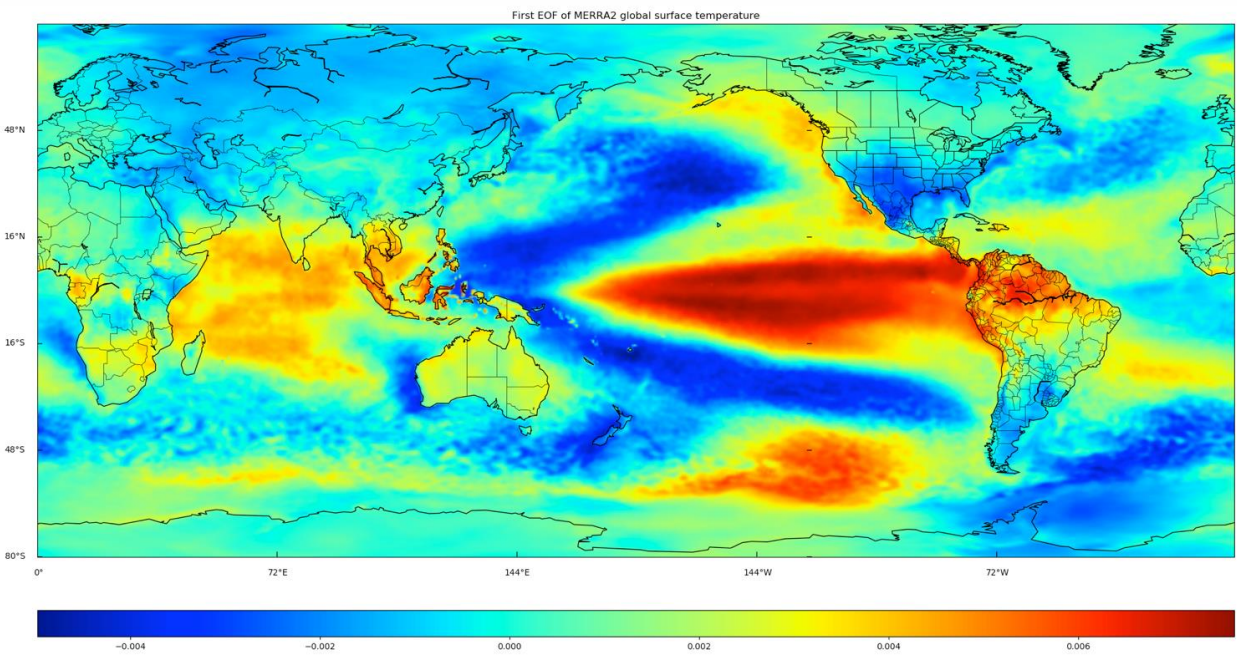

*Figure 5: EDAS Results: MERRA2 Global Surface Temperature, First EOF*

# Visus Streaming Analysis and Visualization

In 2018 the Visus streaming analysis and visualization component has been deployed in a Docker container with three main services exposed: a data streaming server, an on-demand conversion service for NetCDF data and a web viewer that allows interactive visualization of both 2D and 3D datasets directly from the browser. In particular, the on-demand conversion service allows a user to request data from a NetCDF dataset and have it converted into a streamable format (IDX) transparently to the user. Moreover, the service has been enriched with a new metadata library that allows it to copy over all the metadata contained in the NetCDF dataset to the new IDX data format.

In Figure 6 we depict the architecture of the Visus Docker container and the workflow of a user requesting data. The container, based on the file requested by the user, triggers a conversion only if necessary (e.g., the data is not already in the streamable data format). The data will be served directly from the streaming server via HTTP. Together with the web viewer, we developed a javascript API to query data from the streaming server and demonstrated interactive visualization on a browser using large scale climate data.

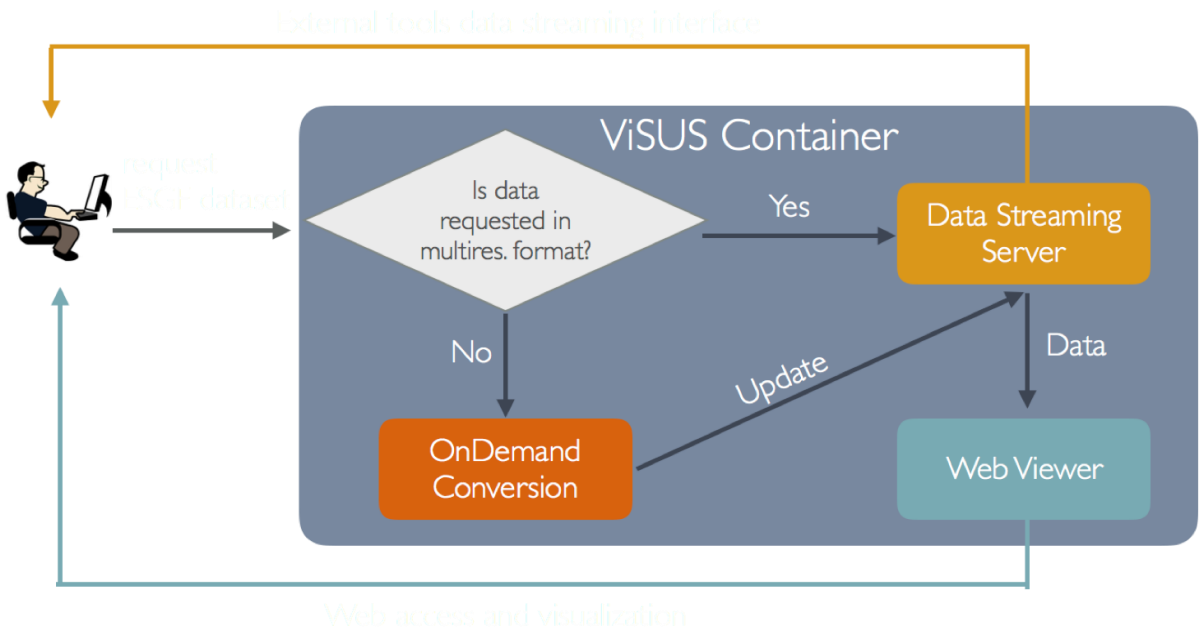

*Figure 6: Visus container architecture*

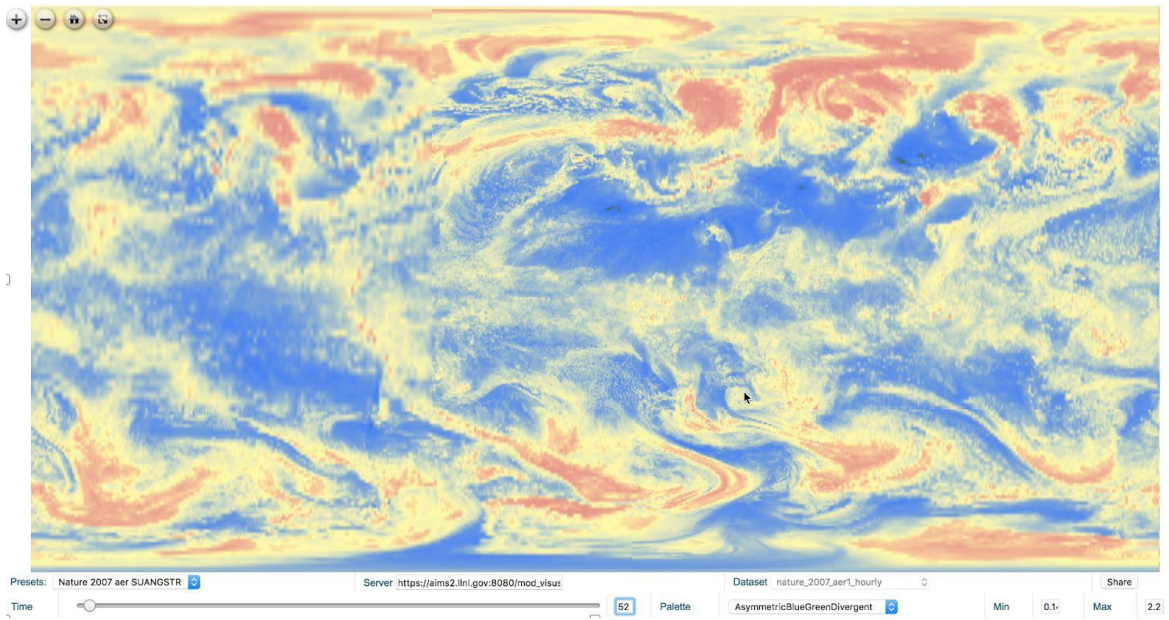
*Figure 7: Visus web viewer*

The webviewer (in Figure 7) allows customization of the visualization through common control settings like: palette, time, slice (for 3d volumes), colormap range, etc. Furthermore it provides the possibility to share a particular view (position and visualization settings) by generating a shareable link.

Another important step to facilitate integration is the deployment of the Visus libraries as a python library available via "pip". This simplifies significantly the distribution of this software and the installation on several platforms. This python library has also been tested through the Jupyter framework creating examples of data exploration and analysis. Also in this case the data can be streamed from the server to the Jupyter notebook client efficiently at different subsampling resolutions (see Figure 8).

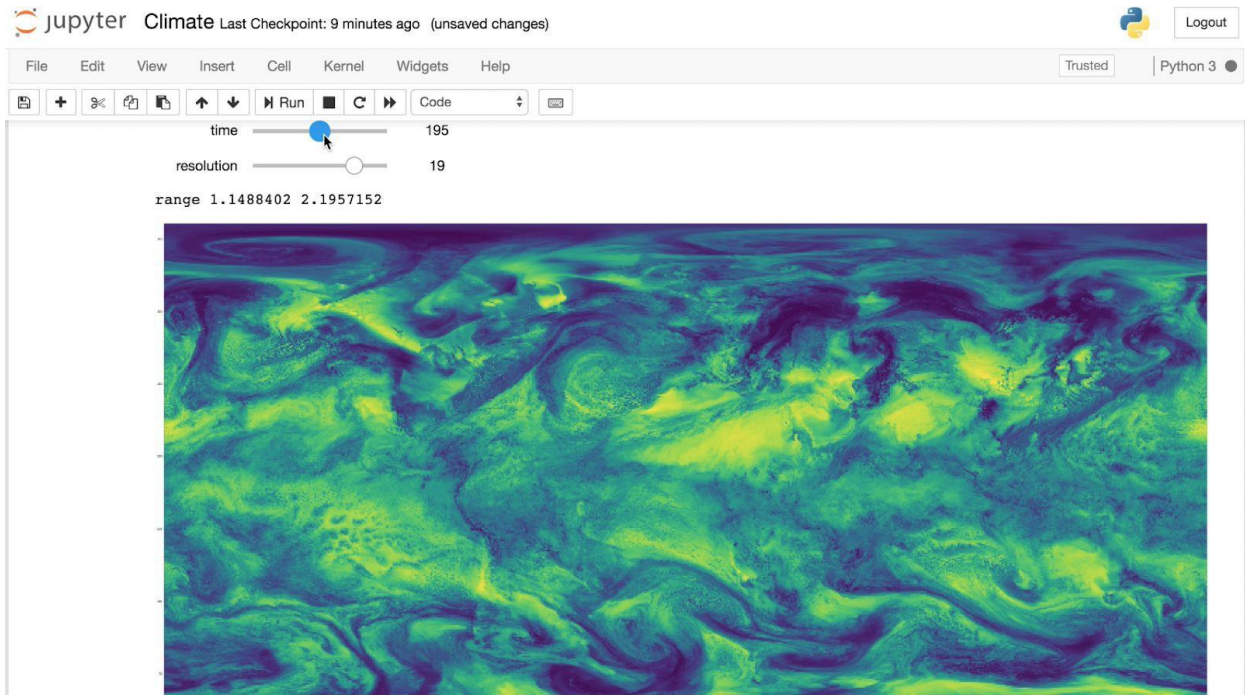

*Figure 8: Jupyter notebook remote data streaming for interactive analysis support*

In 2018 during the development of those features we did some integration tests with the new ESGF containerized architecture. In the next year we plan to finalize this integration and allow the users of the ESGF services to navigate their data interactively directly from the ESGF data portal and also explore the potential of prototyping (or performing) their analysis using data streaming capabilities. Furthermore, improvements to the web interfaces, metadata management and conversion services will facilitate the adoption of an increasing number of users.

## ESGF/Pangeo Testbed

Pangeo is a new community of scientists and software engineers that is focused on providing scalable services for analysis of Earth Science data. The Pangeo infrastructure is based on suite of Python tools, which include Jupyter notebooks for interactive analysis and exploration, Xarray as a high level data model, and Dask for distributing the analysis on all available computing resources. Recently, Pangeo was very successful in demonstrating high performance scalable analysis of climate change data from a variety of sources, including NASA and NOAA remote sensing data, running on an on-premises High Performance Computing Cluster, or on a Cloud cluster managed via Kubernetes.

In the last few months of 2018, the DREAM project collaborated with members of the Pangeo project to explore new architectures for analysis of climate change data held by ESGF. Using the ESGF/Docker software stack, a new ESGF Node was deployed on the Google cloud using the Google Kubernetes Engine (GKE), and populated with sample CMIP6 model data. Then, a Pangeo Jupyter notebook was set up to read and analyze these data via OpenDAP, using a scalable number of Dask workers (see Figure 9). This resulted in a much higher throughput performance than could have been obtained via a traditional, single-threaded application. It also highlighted that the limiting factor in the analysis was reading data from disk via OpenDAP. Therefore, in the next year the DREAM and Pangeo developers intend to experiment with more efficient methods of storing and accessing the data, namely using S3 object store and storing data in a new more suitable compressed format (Zarr).

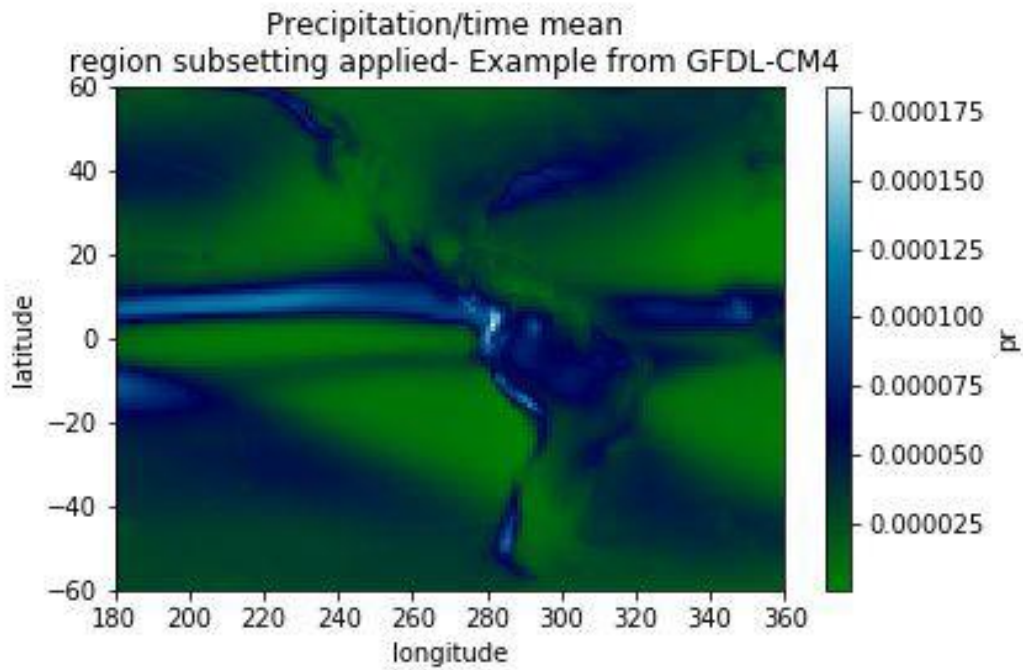

*Figure 9: Sample analysis plots produced using CMIP6 data hosted on the GCP ESGF Node, via the Pangeo data processing infrastructure.*

## ESGF Solr Cloud Index Node

In addition to converting the full ESGF architecture to Docker containers, the DREAM project also embarked on the task of designing, implementing and prototyping the next generation of ESGF Search Services. This task was motivated by some intrinsic limitations in the current ESGF search architecture:
- The high likelihood of inconsistencies between searches initiated at different Nodes.
- The difficulty of scaling the architecture as more nodes join the federation, as the current implementation replicates each others' catalogs.
- The difficulty to upgrade to newer versions of Solr, as this would require a coordinated effort from all Nodes in the federation to prevent long interruptions in replication, and therefore inconsistencies of results.

In response to these limitations, the DREAM project designed a new more scalable architecture for the ESGF Search Services, which was presented at the 8th Annual ESGF Face-To-Face conference in December 2018 (see Figure 10). The proposal is to dispense with replicating catalogs at all search endpoints, but instead to operate a single "ESGF Super-Index" which periodically synchronizes its metadata with each Index Node in the federation, and which is used by all clients (ESGF web portals, scripts, etc.) to obtain results from the full ESGF metadata space.

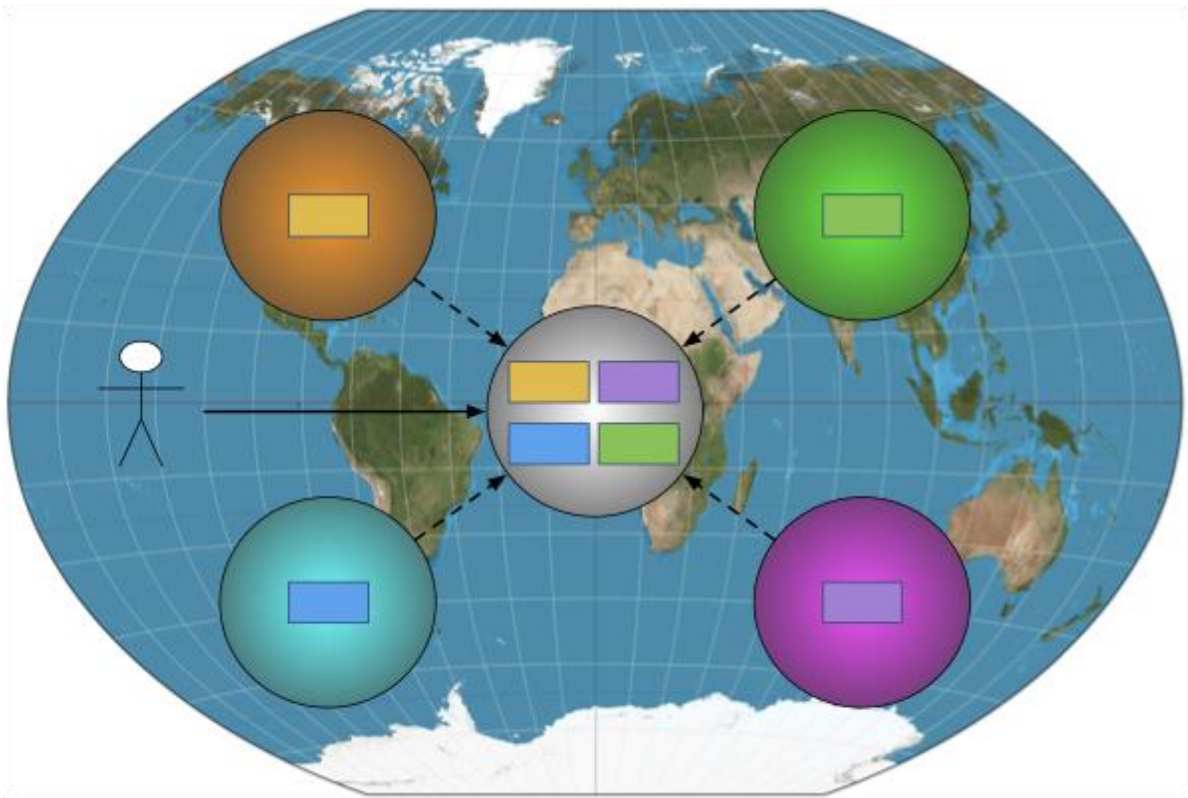

Figure 10: High level concept of the ESGF Super-Index Node.

Internally, the proposed ESGF Super-Index is based on Solr Cloud, a more advanced and scalable implementation of the Solr search engine, and is deployed as a set of Docker containers each running a Solr instance (see Figure 11). The full metadata index is divided into logical shards and physical shard replicas, which are deployed on different containers for resiliency and fault tolerance (see Figure 12). A set of Python processes was developed to execute the initial harvesting of metadata from each ESGF Index Node to the Super-Index, and to then execute periodic synchronization.

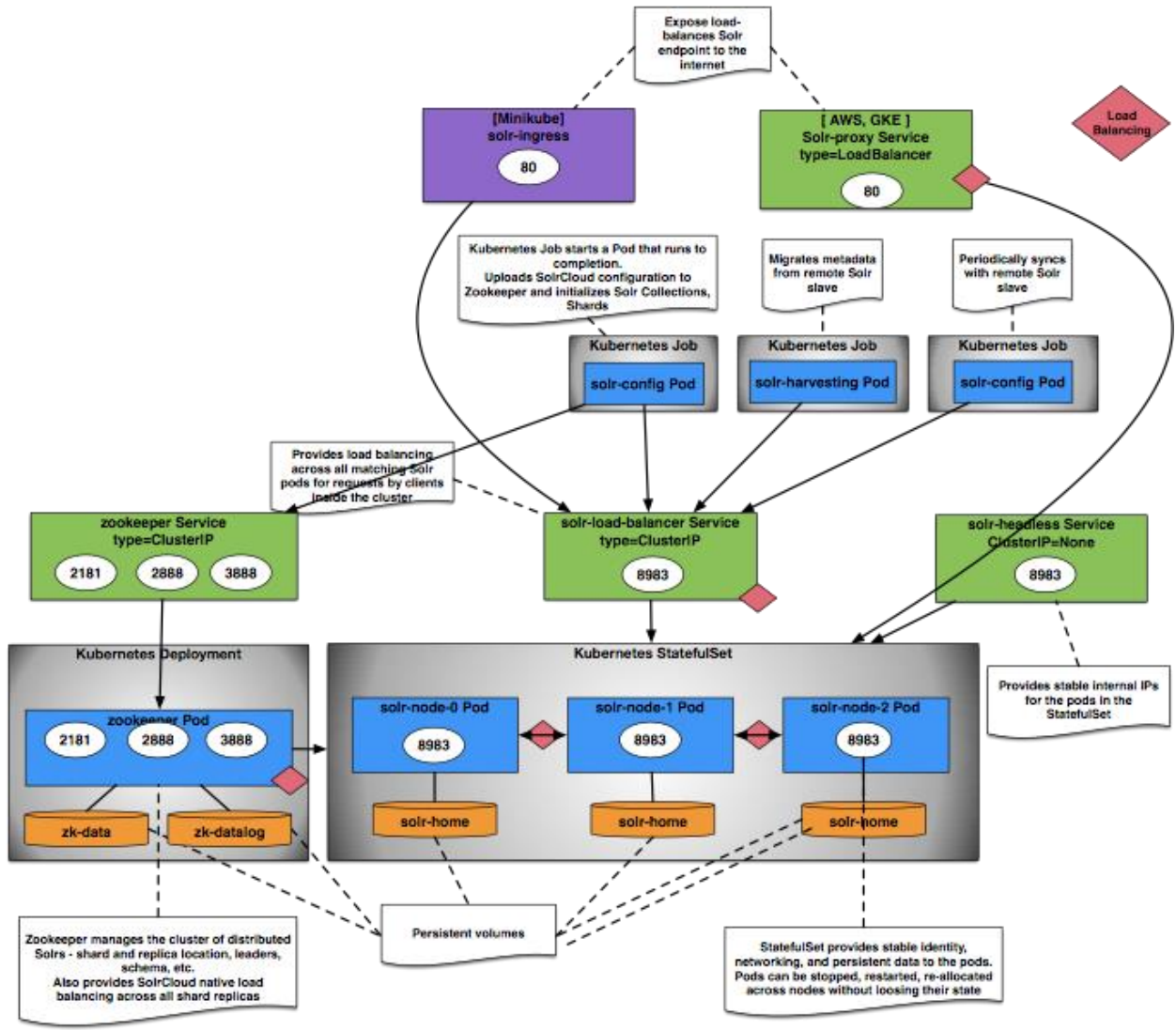

*Figure 11: Internal architecture of the ESGF Super-Index Node as a set of Docker containers deployed as different kinds of Kubernetes objects.*

In order to prove the viability of this approach, a prototype Super-Index was deployed on the Amazon cloud, using a rather small set of EC2 instances (three "t2.medium" servers with 2 cores and 4GB of memory each). This small cluster actually proved to be sufficient to faithfully track the global ESGF metadata holdings over the course of several months.

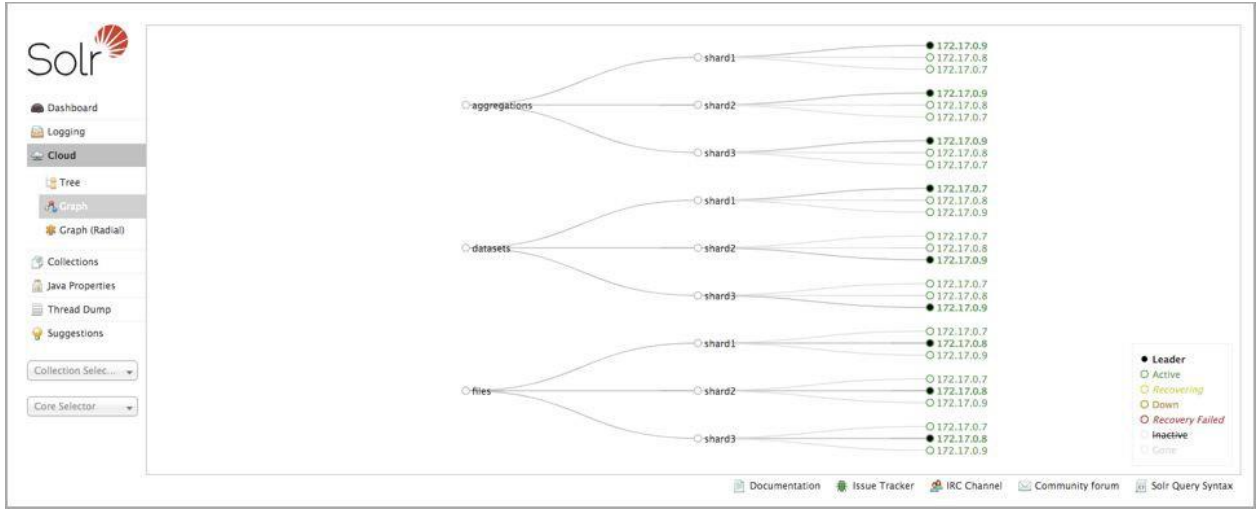

*Figure 12: Prototype deployment of the ESGF Super-Index on AWS, showing each metadata collection ("Datasets", "Files", "Aggregations") as split into 3 shards, each with 3 replicas.*

In the next year, pending final approval by the ESGF Executive Committee, the DREAM project plans to deployed this architecture as an operational service running on the Amazon or Google cloud, and to promote this deployment as the single endpoint that should be used by all ESGF search clients.

# Biology Testbed

The completed biology testbed focuses on publishing and accessing several datasets from sample genomics applications. The testbed assumes the installation of an ESGF datanode. We augment this installation with the configuration of a project for data publication and the DREAM data service used for data access. We make use of several helper scripts to publish data. The service (prototype) shows a proof of concept of remote .fasta file query. A technical report that describes the data and entire process from project definition through access is close to completion and will be made available on the DREAM website.

# Python-based Build, Installation, and Deployment

The original ESGF installer, based on bash scripting, contained a monolithic file with most of the code, and several smaller sub-modules deployed via project-maintained mirrors at several participant sites. The deployment model evolved over a decade under several different developers, and has turned out to be unmaintainable. The disorganized and monolithic base of code has stymied development and cost significant maintenance overhead.

Our Python deployment reorganized the components. All sub-modules reside in a single code repository and are designed around node functionality, either as part of the node installation (different type of node), node control, (eg. start, stop, status), or node utility (eg. shard, certificate management). Considerable effort was needed to understand the operations of the original routines in bash, and then implement comparable routines in Python

While we have completed our effort for the DREAM project, the work on the Python installation has informed our approach to the next effort. Ansible (https://docs.ansible.com/) has become a remarkably mature platform for system software deployment, which fits the requirements for ESGF. Rather than continue down the path and re-implement the remaining outstanding features within python to improve resiliency, we will opt to use Ansible given its success as an established deployment platform.

# Future Milestones

- **ESGF Node Containerized architecture**
    - Release v2.0 of the software stack, to be feature complete and to be on-par with the latest service versions installed by the classic ESGF installer.
    - Enable deployment of different topologies using the shared pool of ESGF/Docker images – i.e. the capability to deploy a full ESGF node, or just an Index node, a Data node, etc.
- **Solr Cloud ESGF Super Index**
    - Deploy the newly designed architecture on AWS or GCP, harvest all metadata holdings in the federation, and operate as the preferred endpoint to be used by all ESGF search clients.
- **ESGF/Pangeo testbed**
    - Further develop the ESGF/Pangeo testbed by converting a sub-sample of CMIP6 data to Zarr format and store it on S3 on Google Cloud, then use the Pangeo infrastructure to more efficiently read and analyze the data.
- **Visus**
    - Integration of streaming visualization and on-demand conversion services with ESGF-CoG data portal. The related container will be part of the ESGF containerized architecture.
- **ESGF Compute Node**
    - Release v2.0 of the EDAS analytics server featuring ESGF integration, compliance, and certification as per the ESGF Compute Challenge.
    - Extend EDAS capabilities with the Stratus Microservices Integration Framework (under development) to facilitate the assimilation of external analytic services into EDAS workflows.

# Products

- Software:
    - ESGF/Docker source code repository: https://github.com/ESGF/esgf-docker
    - ESGF/Ansible installer: https://github.com/ESGF/esgf-ansible
    - ESGF/Visus integration: https://github.com/ESGF/esgf-dream-data-service
- Talks:
    - "The State of the Earth System Grid Federation", 8th Annual ESGF Face-To-Face Workshop, Luca Cinquini (NASA/JPL), Washington DC, December 2018
    - "Proposal for Next Generation ESGF Search Services", 8th Annual ESGF Face-To-Face Workshop, Luca Cinquini (NASA/JPL), Washington DC, December 2018
    - "Pangeo: a flexible climate analytics infrastructure for climate data on ESGF", 8th Annual ESGF Face-To-Face Workshop, Ryan Abernathy (Columbia Univ.), Washington DC, December 2018
    - "DREAM Final Report", NGNS PI meeting, Luca Cinquini (NASA/JPL), Rockville, MD, September 2018
    - "The Magic of Dreams", MAGIC web-hosted seminar, Luca Cinquini (NASA/JPL), June 2018
- Posters
    - "The Earth Data Analytic Services (EDAS) Framework", 2018 Fall Meeting, Thomas Maxwell, Washington DC, December 2018

# Participants and Other Collaborating Organizations

- Participant: Ames, Sasha (LLNL)
    - Project role: LLNL Senior Personnel, acting LLNL PI
    - Person months worked: 1
    - Funding support (if other than this award): DOE/ESGF, DOE/PCMDI
    - Contribution to the project: biology datasets prototype publication and site configuration
    - International collaboration: ESGF international partners
    - International travel: none in 2018
- Participant: Balaji, V. (Princeton)
    - Project role: Princeton Principal Investigator
    - Person months worked: 1
    - Funding support (if other than this award): NOAA
    - Contribution to the Project: PI, project design and technical oversight
    - International collaboration: none

- International travel: none in 2018
- Participant: Christensen, Cameron (University of Utah)
  - Project role: University of Utah Senior Software Developer
  - Person months worked: 1
  - Funding support (if other than this award): DOE, NSF
  - Contribution to the Project: development of the data streaming infrastructure for data analysis and visualization
  - International collaboration: ESGF international partners
  - International travel: none in 2018
- Participant: Cinquini, Luca (JPL)
  - Project role: JPL Principal Investigator
  - Person months worked: 9 in 2018
  - Funding support (if other than this award): NASA, DOE
  - Contribution to the Project: overall system architecture, containerization, application to climate science
  - International collaboration: ESGF international partners
  - International travel: none in 2018
- Participant: Doutriaux, Charles (LLNL)
  - Project role: LLNL Co-I
  - Person months worked: 1
  - Funding support (if other than this award): DOE/ESGF, DOE/ACME, DOE/PCMDI
  - Contribution to the Project: co-leading the ESGF-CWT working group
  - International collaboration: ESGF, UV-CDAT, ACME national and international partnerships
  - International travel: none in 2018
- Participant: Duffy, Dan (GSFC)
  - Project role: GSFC Principal Investigator
  - Person months worked: 1
  - Funding support (if other than this award): NASA
  - Contribution to the Project: co-leading the ESGF-CWT working group
  - International collaboration: none
  - International travel: none in 2018
- Participant: Ferraro, Robert (JPL)
  - Project role: JPL Co-Investigator
  - Person months worked: 0.5
  - Funding support (if other than this award): NASA
  - Contribution to the Project: user community requirements, ESGF Executive Committee, interoperability with NASA DAACs
  - International collaboration: ESGF international partners
  - International travel: none
- Participant: Greguska, Frank (JPL)
  - Project role: Senior Software Developer

- ○ Person months worked: 3 in 2018
- ○ Funding support (if other than this award): NASA
- ○ Contribution to the Project: Cloud deployments, containerization, hydrology testbed
- ○ International collaboration: Science & Technology Facilities Council, ESGF international partners
- ○ International travel: None
● Participant: Hill, William (LLNL)
  - ○ Project role: Senior Software Developer
  - ○ Person months worked: 6
  - ○ Funding support (if other than this award): DOE
  - ○ Contribution to the Project: contributing to the ESGF installation efforts by leading the rewrite of the installation
  - ○ International collaboration: ESGF installation team
  - ○ International travel: none
● Participant: Boutté, Jason (LLNL)
  - ○ Project role: Software Developer
  - ○ Person months worked: 8
  - ○ Funding support (if other than this award): DOE
  - ○ Contribution to the Project: development of the LLNL server-side analytics within the ESGF-CWT working group
  - ○ International collaboration: ESGF Compute Working Tea,
  - ○ International travel: none
● Participant: Maxwell, Thomas (GSFC)
  - ○ Project role: Senior Software Developer
  - ○ Person months worked: 12
  - ○ Funding support (if other than this award): NASA
  - ○ Contribution to the Project: development of server-side analytics (EDAS) within the ESGF-CWT working group
  - ○ International collaboration: none
  - ○ International travel: none in 2018
● Participant: Nikonov, Sergey (Princeton)
  - ○ Project role: Princeton Technical Specialist
  - ○ Person months worked: 7.25
  - ○ Funding support (if other than this award): NOAA
  - ○ Contribution to the Project: data volume estimates for DREAM; participation in ESGF WTs; design and deployment of containerized climate analytics
  - ○ International collaboration: none
  - ○ International travel: none in 2018
● Participant: Pascucci, Valerio (University of Utah)
  - ○ Project role: University of Utah Principal Investigator
  - ○ Person months worked: 0.25
  - ○ Funding support (if other than this award): DOE, NSF

- - Contribution to the Project: system architecture of the data streaming infrastructure
    - International collaboration: ESGF international partners
    - International travel: none in 2018
- Participant: Petruzza, Steve (University of Utah)
  - Project role: University of Utah PostDoctoral Fellow
  - Person months worked: 2
  - Funding support (if other than this award): DOE, NSF
  - Contribution to the Project: containerization, web and python interfaces
  - International collaboration: ESGF international partners
  - International travel: none in 2018
- Participant: Vahlenkamp, Hans (UCAR, Princeton subcontract)
  - Project role: Princeton Investigator
  - Person months worked: 6
  - Funding support (if other than this award): NOAA
  - Contribution to the Project: climate modeling workflow and integration with ESGF Docker services
  - International collaboration: none
  - International travel: none in 2018

# Impact

The DREAM project is having a critical impact in determining the future strategic direction of the Earth System Grid Federation, and more generally, in influencing the analysis of climate data by the international community, due to the following accomplishments:
- DREAM developed 2 new procedures for installing and configuring an ESGF node, one based on Docker containers, the other on Ansible scripting. Both these procedures are being adopted by ESGF as an immediate replacement for the obsolete and unmaintainable shell-based installer, thus greatly improving the easiness and flexibility of installing the ESGF software stack, and consequently the general adoption of ESGF by climate centers across the globe.
- The Solr-Cloud architecture, implemented as Docker containers, is under consideration by ESGF for replacing the current search services at each local node. The goal is to make this a central, high performance and fault tolerance endpoint for search and discovery of climate datasets everywhere. Having a single search endpoint will also greatly facilitate upgrading the search services within the federation, as well as responding to security threats and vulnerabilities.
- The ESGF Compute Node stack is about to be integrated with the rest of the ESGF deployment, and will therefore accomplish the long sought goal of "moving the computation to the data". When deployed at multiple nodes throughout the federation, it should open new possibilities for conducting scientific research on climate change.

- The ESGF/Pangeo prototype testbed has sparked considerable interest, and is leading to exciting discussions within the WGCM (the "World Climate Research Programme") community for scaling the analysis of massive datasets stored on the Cloud through a Python-based infrastructure. These ideas will likely be prototyped using storage donated by Google (and perhaps Amazon) to host a significant portion of the CMIP6 model output.

# Problems

- No significant problems were encountered at any of the institutions participating in the DREAM project.

## Acknowledgments

The results of this research use in part IP that is included in the University of Utah Patents number: 9442960 "High performance data layout and processing" and number: 8836714 "Rapid, interactive editing of massive imagery data". The University of Utah granted exclusive rights to ViSUS LLC owned by Valerio Pascucci.